\documentclass[12pt,letterpaper]{article}
\usepackage[latin1]{inputenc}
\usepackage{amsmath}
\usepackage{amsfonts}
\usepackage{amssymb}
\usepackage{graphicx,color}

\newcommand{\bet}{\beta}

\newcommand{\al}{\alpha}
\newcommand{\ga}{\gamma}
\newcommand{\de}{\delta}
\newcommand{\ep}{\epsilon}
\newcommand{\la}{\lambda}
\newcommand{\om}{\omega}
\newcommand{\si}{\sigma}
\newcommand{\rh}{\rho}
\newcommand{\be}{\begin{equation}}
\newcommand{\ee}{\end{equation}}
\newcommand{\bea}{\begin{eqnarray}}
\newcommand{\eea}{\end{eqnarray}}

\begin{document}

\thispagestyle{empty}

\setcounter{page}{0}

\mbox{}
%\vspace{-20mm}

\begin{center} {\bf \Large  Gravitational Electric-Magnetic Duality, Gauge Invariance and Twisted Self-Duality}\footnote{Invited contribution to the  J Phys A special volume on ``Higher Spin Theories and AdS/CFT" edited by Matthias Gaberdiel and Misha Vasiliev.}

\vspace{1.6cm}

Claudio Bunster$^{1,2}$,  Marc Henneaux$^{1,3}$ and Sergio H\"ortner$^3$

\footnotesize
\vspace{.6 cm}

${}^1${\em Centro de Estudios Cient\'{\i}ficos (CECs), Casilla 1469, Valdivia, Chile}

\vspace{.1cm}

${}^2${\em Universidad Andr\'es Bello, Av. Rep\'ublica 440, Santiago, Chile}

\vspace{.1cm}

${}^3${\em Universit\'e Libre de Bruxelles and International Solvay Institutes, ULB-Campus Plaine CP231, B-1050 Brussels, Belgium} \\

\vspace {15mm}

\end{center}
\centerline{\bf Abstract}
\vspace{.6cm}
The concept of electric-magnetic duality can be extended to linearized gravity. It has  indeed been established that in four dimensions, the Pauli-Fierz action (quadratic part of the Einstein-Hilbert action) can be cast in a form that is manifestly invariant under duality rotations in the internal $2$-plane of the spacetime curvature and its dual.  In order to achieve this manifestly duality-invariant form, it is necessary to introduce two ``prepotentials", which form a duality multiplet.  These prepotentials enjoy interesting gauge invariance symmetries, which are, for each,   linearized diffeomorphisms and linearized Weyl rescalings.  The purpose of this note is twofold: (i) To rewrite the manifestly-duality invariant action obtained in previous work  in a way that makes its gauge invariances also manifest. (ii) To explicitly show that the equations of motion derived from that action can be interpreted as twisted self-duality conditions on the curvature tensors of the two metrics obtained from the two prepotentials.

\vspace{.8cm}
\noindent

\newpage

\section{Introduction}
\setcounter{equation}{0}
\subsection{Einstein equations as twisted self-duality conditions}

There exists an interesting formulation of electromagnetism in four dimensions in which one introduces two potentials.  One potential  is the usual ``electric potential" $1$-form A and the other is the  ``magnetic  potential" $1$-form  B. If one demands that the corresponding field strengths (curvature $2$-forms) be the dual of each other one obtains the Maxwell equations. This requirement is called ``twisted self-duality" \cite{Cremmer:1998px}.  The term ``twisted"  is introduced because the  forms are not self-dual but are, rather,  dual to each other. If both curvature forms are grouped into a two-component colum, then that colum is related to its dual by an off-diagonal ``twist matrix".

The formulation can be extended to free $p$-form gauge fields  in any number of dimensions and one may also include certain types of couplings  \cite{Cremmer:1998px}.  In all cases, the twisted self-dual equations of motion can be derived from a variational principle in which the two potentials are on an equal footing \cite{Deser:1976iy,Henneaux:1988gg,ScSe,Tonin,DGHT,Bunster:2011aw,Bunster:2011qp}. The ``twisted two-potential action" can be systematically obtained from the Hamiltonian formulation of the theory by solving the constraints  \cite{Deser:1976iy,Bunster:2011qp}, a method that indicates that the two potentials can be viewed as forming a generalized canonically conjugate pair.

An interesting feature of the twisted two-potential action emerges in the special dimension where the two potentials are exterior forms of the same rank $p$. This occurs when $D= 2p+2$.  The cases $p$ odd and $p$ even exhibit different properties \cite{DGHT}. When $p$ is odd ($D=0$ modulo $4$), the Hodge operator squares to minus one and the two-potential action exhibits a continuous  symmetry, which is the group of $SO(2)$ rotations in the internal plane spanned by the two field-strengths.  The simplest case, $p=1$, is just the standard ``electric-magnetic" duality, and for this reason, this group of $SO(2)$ rotations is called ``electric-magnetic" duality also when $p \not=1$.  When $p$ is even ($D=2$ modulo $4$), the Hodge operator squares to one and one may ``untwist the twisting", i.e. diagonalize  the twisting matrix over the reals and split the fields into ``chiral" and ``anti-chiral"  components, respectively corresponding to the eigenvalues $1$ and $-1$.

The twisted self-duality formulation of the equations of motion can be extended to linearized gravity.  In this case, instead of exterior forms of rank $p$ and $D-p-2$, dualization involves the standard  rank-2 symmetric tensor corresponding to the usual description of the graviton and a dual mixed tensor field with two columns, one of length $D-3$ and one of unit length \cite{Curtright:1980yk,Hull:2001iu}.

We shall consider here explicitly the case $D=4$, deferring the discussion of the general case to a subsequent publication \cite{InPrep}.  In $D=4$ the two dual fields are both rank-2 symmetric tensors.  

The twisted self-dual formulation proceeds then as follows.
Let $h_{\mu \nu}$ and $f_{\mu \nu}$ be two rank-2 symmetric tensors, and $R_{\lambda \mu \rho \sigma}$, $S_{\lambda \mu \rho \sigma}$ their linearized curvatures,
\begin{eqnarray}
R_{\lambda \mu \rho \sigma} &=& - \frac{1}{2} \left(\partial_\lambda \partial_\rho h_{\mu \sigma} -  \partial_\mu \partial_\rho h_{\lambda \sigma} - \partial_\lambda \partial_\sigma h_{\mu \rho} + \partial_\mu \partial_\sigma h_{\lambda \rho}\right), \nonumber \\
S_{\lambda \mu \rho \sigma} &=& - \frac{1}{2} \left(\partial_\lambda \partial_\rho f_{\mu \sigma} -  \partial_\mu \partial_\rho f_{\lambda \sigma} - \partial_\lambda \partial_\sigma f_{\mu \rho} + \partial_\mu \partial_\sigma f_{\lambda \rho}\right). \nonumber
\end{eqnarray}
The dual curvatures are defined through
\begin{eqnarray}
^*R_{\lambda \mu \rho \sigma} &=& \frac{1}{2} \epsilon_{\lambda \mu \alpha \beta} R^{\alpha \beta}_{\; \; \; \; \; \;  \rho \sigma}, \nonumber \\
^*S_{\lambda \mu \rho \sigma} &=& \frac{1}{2} \epsilon_{\lambda \mu \alpha \beta} S^{\alpha \beta}_{\; \; \; \; \; \;  \rho \sigma}. \nonumber 
\end{eqnarray}
Our conventions are $\epsilon_{0123} = 1 = - \epsilon^{0123}$, so that in particular $\epsilon_{0ijk} =  \epsilon_{ijk}$.

The ``twisted self-duality conditions", which express that $R$ is the dual of $S$ (we drop indices)
$$
R = \; \!^*S, \; \; \; S = \;  -\;  \!^*R,
$$
or,
\begin{equation}
{\mathfrak R} = - {\mathcal S}  \, ^*\hspace{-.02cm}{\mathfrak R}, \label{122}
\end{equation}
with
\begin{equation}
{\mathfrak R} = 
 \begin{pmatrix} R\\ S\\ \end{pmatrix}, \; \; \; {\mathcal S}  = \begin{pmatrix} 0&-1 \\ 1 & 0 \end{pmatrix}  ,\label{123}
\end{equation}
%\vspace{.3cm}
imply that $h_{\mu \nu}$ and $f_{\mu \nu}$ are both solutions of the linearized Einstein equations,
$$
R_{\mu \nu} = 0, \; \; \;  S_{\mu \nu} = 0.
$$
This is because the cyclic identity for $S$ (respectively, for $R$) implies that the Ricci of $R$ (respectively, of $S$) vanishes.

Conversely, if $h_{\mu \nu}$ is a solution of the Einstein equations,  the dual $^*R_{\lambda \mu \rho \sigma}$ of its Riemann tensor obeys the cyclic identity and so can be viewed as the Riemann tensor $S_{\lambda \mu \rho \sigma}$ of a ``dual" symmetric tensor $f_{\mu \nu}$ which also obeys the Einstein equations.  To avoid proliferation of symbols, we shall in the sequel write $R_{\lambda \mu \rho \sigma}[f]$ instead of $S_{\lambda \mu \rho \sigma}$ and $R_{\lambda \mu \rho \sigma}[h]$ for the Riemann tensor of $h_{\mu \nu}$.

\subsection{Duality symmetry}

Because the original graviton field $h_{\mu \nu}$ and its dual $f_{\mu \nu}$ are both rank-2 symmetric tensors, $D=4$ is a special dimension in which ``electric magnetic-duality" emerges.  Just as for electromagnetism, this $SO(2)$ continuous symmetry is given by the group of rotations in the internal plane spanned by the curvature and its dual. 

It was shown in \cite{Henneaux:2004jw} that gravitational electric-magnetic duality was not just a symmetry of the equations of motion, but again, also a symmetry of the action{\footnote{This implies the existence of a conserved Noether charge, which acts as the generator of duality rotations through the Poisson bracket (classically)
or the commutator (quantum-mechanically).  The existence of a Noether generator would not hold if duality was a mere equations-of-motion symmetry.}. The analysis was extended to spin-$2$ in de Sitter space in \cite{Julia:2005ze}{\footnote{The method of   \cite{Julia:2005ze} considers a flat slicing of de Sitter space and can be straightforwardly extended to any space-time that admits such a slicing as does locally anti-de Sitter space. It would be of great interest to go beyond flat slicings  but this question does not fall within the scope of this paper. } (see also \cite{Nieto:1999pn} and \cite{Deser:2004xt} for different approaches).

The derivation of \cite{Henneaux:2004jw} relies on the canonical formulation of the theory and introduces two conjugate  ``prepotentials", one for the spatial components $h_{ij}$ of the metric and one of their conjugate momenta $\pi^{ij}$.  Explicitly, these are defined through
\bea
h_{ij} &=& \ep_{irs} \partial^r \Phi^{s}_{\; \; j} + \ep_{jrs} \partial^r \Phi^{s}_{\; \; i} + \partial_i u_j + \partial_j u_i \label{hPhi0},\\
\pi^{ij} &=& \ep^{ipq} \ep^{jrs} \partial_p \partial_r P_{qs} \label{piP}
\eea
(the vector $u_i$ can also be thought of as a prepotential but it drops out from the theory so that we shall not put emphasis on it).   When reformulated in terms of the prepotentials, duality symmetry  simply amounts to $SO(2)$ rotations in the internal plane of the prepotentials.  

The action reads explicitly \cite{Henneaux:2004jw}
\be
S[Z_a^{\; mn}] = \int dt \left[ \int d^3x \, \ep^{ab} \ep^{mrs} \left(\partial^p \partial^q \partial_r Z_{aps} - \triangle \partial_r Z_{a\;   s}^{\;q}\right) \dot{Z}_{bqm} - H \right] \label{Action0}
\ee
($a, b = 1,2$) where $(Z^a_{ij}) \equiv (P_{ij},\Phi_{ij})$ are the prepotentials and $\triangle$ the Laplacian.
 Here the Hamiltonian is given by\footnote{The reader might be surprised to find a non-vanishing Hamiltonian for the linearized spin-two theory, while the Hamiltonian for the full Einstein theory, generally thought to reduce to a combination of constraints, would seem to vanish.  The point is that the Hamiltonian of Einstein gravity does {\em not} vanish in the asymptotically flat context, where it is given by a  combination of the constraints {\em plus} a non-vanishing surface integral. Using the constraints of the full Einstein theory, one may convert the surface integral into a volume piece which, to quadratic order, is given by the volume integral of the energy density of the spin-2 field. }
\bea
H &=& \int d^3x\,  \de^{ab} \left( \triangle Z_{aij} \, \triangle Z_b^{\; ij} + \frac{1}{2} \partial^k \partial^m Z_{akm} \partial^q\partial^n Z_{bqn} + \partial^k \partial^m Z_{akm} \triangle Z_{b}\right) \nonumber \\
&& +  \int d^3x \, \de^{ab} \left(-2 \partial_m \partial_i Z_{a}^{\; ij} \partial^m\partial^k Z_{bkj} - \frac{1}{2} \triangle Z_{a} \triangle Z_{b}\right).
\eea
In these $SO(2)$ vector notations, duality invariance is manifest.

As as shown in \cite{Henneaux:2004jw}, the manifestly duality invariant action (\ref{Action0}) is also invariant under gauge symmetries, which are  linearized diffeomorphisms and Weyl rescalings of the prepotentials,
\be
\delta Z^a_{\; ij} = \partial_i \xi^a_{\; j} + \partial_j \xi^a_{\; i} + 2 \ep^a \de_{ij} . \label{GaugeTransf}
\ee
These gauge invariances can be checked by direct substitution.

\subsection{Purposes of this paper}

The purposes of this paper are twofold.

First,  the gauge symmetries of the action (\ref{Action0}) are not entirely obvious in the form the action has been written.
One would like to make gauge invariance manifest, and to understand better the structure of the action, by using the appropriate tensor calculus.  This is an interesting task because,
although possessing the gauge symmetries (\ref{GaugeTransf}) of linearized conformal gravity, the action (\ref{Action0}) is not the  action of that theory: it involves two independent symmetric tensors, it is first order, it is not manifestly Lorentz-invariant and, as we have explained,  it is equivalent to the standard Pauli-Fierz action. 

Second, while duality invariance was established in \cite{Henneaux:2004jw}, it was not explicitly verified there that the equations of motion following from the action (\ref{Action0}) were indeed the twisted self-duality conditions (\ref{122}).  The spatial components $f_{ij}$ of the dual metric are defined in terms of $P_{ij}$ as the $h_{ij}$'s are defined in terms of $\Phi_{ij}$, 
\be
f_{ij} = \ep_{irs} \partial^r P^{s}_{\; \; j} + \ep_{jrs} \partial^r P^{s}_{\; \; i} + \partial_i v_j + \partial_j v_i ,\label{fijP}
\ee
where $v_i$ is some other prepotential reflecting the diffeomorphism invariance of $f_{ij}$, which drops out from the curvature.  While the fact that the equations following from (\ref{Action0}) are indeed equivalent to the twisted self-duality equations (\ref{122}) is an implicit consequence of the analysis of \cite{Henneaux:2004jw}, it is not devoid of interest to prove it directly.  Indeed, the equations following from (\ref{Action0}) are of first order in the time derivatives of the prepotentials and hence also of the metrics $h_{ij}$ and $f_{ij}$, while the the twisted self-duality equations (\ref{122}) involve also second order time derivatives.  

Establishing this equivalence is furthermore useful in other space-time dimensions \cite{InPrep} where twisted self-duality is all there is (there is no $SO(2)$-duality since the metric and its dual are of different tensor types).

Our paper is organized as follows: In the next section, Section \ref{Twisted}, we reformulate the twisted self-duality condition in a form adapted to the $(3+1)$-splitting of spacetime.  Then, in section \ref{GaugeInvAction1}, we rewrite the action in a manner that makes its gauge invariances manifest. We also express the equations of motion following from (\ref{Action0}) in terms of the appropriate tensors invariant under linearized diffeomorphisms and Weyl rescalings. In Section \ref{HETwistedSD}, we verify that these equations of motion are indeed the twisted self-duality conditions obtained in Section \ref{Twisted}. The last section, Section \ref{Conclusions} is devoted to concluding comments. Four appendices (Appendices \ref{CottonTensor}, \ref{Case3D},  \ref{TensorsLineGrav} and \ref{UsefulForm}) recall for completeness some concepts from conformal geometry and apply them to the case at hand.  Finally, Appendix \ref{H1EOMH2EOM}  further analyses the equations of motion for the prepotentials.

\section{$3+1$-Form of the Twisted Self-Duality Conditions}
\label{Twisted}
\setcounter{equation}{0}
\subsection{Twisted self-duality conditions revisited}
There is some redundancy in the covariant form  (\ref{122}) of the twisted self-duality conditions.
To display this feature, it is convenient to introduce the electric and magnetic components of the Riemann tensor defined as follows,
\begin{eqnarray}
&&  {\mathcal E}_{mn} = R_{mn} , \nonumber \\
&& {\mathcal B}_{mn} = - \frac{1}{2} \epsilon_{npq} R_{0m}^{\; \; \; \; \; pq}, \nonumber 
\end{eqnarray}
where $R_{mn}$ denotes the three-dimensional Ricci tensor built out of the spatial components $h_{mn}$ of $h_{\mu \nu}$.  The magnetic field contains one time derivative of the metric, the electric field contains  only spatial derivatives.  Note however that  $R_{0m0n} = - \; ^{(4)}R_{mn} + R_{mn}$ and thus on-shell $ {\mathcal E}_{mn} = R_{0m0n}$.  From now on, we shall affect a subscript $(4)$ to space-time objects when confusion with the corresponding three dimensional object could arise\footnote{A space-time for which the magnetic (respectively electric) field vanishes is called electric (respectively, magnetic).  Thus the (linearized) Schwarzschild solution, and more generally, static solutions, are electric.}.
Note also that in three dimensions, the Riemann tensor is completely determined by the Ricci tensor, so that $ {\mathcal E}_{mn}$ is equivalent to $R_{mnpq}$. 

It follows from the twisted self-duality conditions that,
\begin{eqnarray}
&& {\mathcal B}_{mn}[h] = {\mathcal E}_{mn}[f] , \nonumber \\
&& {\mathcal B}_{mn}[f] = - {\mathcal E}_{mn}[h] ,\nonumber 
\end{eqnarray}
or in matrix notation,
\begin{equation}
 \begin{pmatrix} \mathcal{E}_{mn}[h]\\ \mathcal{E}_{mn}[f]\\ \end{pmatrix} = {\mathcal S}  \, \begin{pmatrix} \mathcal{B}_{mn}[h]\\ \mathcal{B}_{mn}[f]\\ \end{pmatrix} .\label{EBDuality}
\end{equation}

We claim that the twisted self-duality conditions (\ref{EBDuality}) are completely equivalent to (\ref{122}) and in particular imply all of the Einstein equations.  This is a non trivial statement because the conditions (\ref{EBDuality}) are only a subset of (\ref{122}), namely the subset that contains no second time derivatives. 

To establish the claim, we proceed in two steps.

\subsection{Step 1: Constraint Equations}

The constraint equations follow from (\ref{EBDuality}).  \\ {\bf Proof:} It follows from the twisted self-duality relations (\ref{EBDuality}) that the magnetic field is symmetric (while it is not for an arbitrary metric), since the electric field is.  This implies $R_{0i}[h] = 0$, $R_{0i}[f] = 0$. Equivalently, $G_{0i}[h] = G_{0i}[f]= 0$.

Similarly, it follows from the twisted self-duality relations (\ref{EBDuality}) that the electric field is traceless (while it is not for an arbitrary metric), since the magnetic field is.  This implies that the three-dimensional curvature scalars $R[h] = 0 = R[f]$ both vanish. For the linearized theory these are the Hamiltonian constraints $G_{00}[h] = 0 = G_{00}[f]$ for $h_{\mu \nu}$ and $f_{\mu \nu}$.

\vspace{.2cm}
The next step is to establish the dynamical Einstein equations.  This is a bit harder because these involve two time derivatives of the metric, so that one needs to differentiate the self-duality conditions with respect to time.  But this will lead to third order differential equations and so, one can only hope to get the dynamical Einstein equations differentiated once, but without loss of information.  This turns out to be the case.  This is in sharp contrast with the electromagnetic situation, where the twisted self-duality conditions are first-order differential equations, while the Maxwell equations are of second order, so that by differentiating once the twisted self-duality condition, one can derive the Maxwell equations in their standard form .

\subsection{Step 2: Dynamical Equations}

The dynamical Einstein equations also follow from (\ref{EBDuality}).  \\ {\bf Proof:}  We start by computing $\partial_0 R_{i0mj}[h]$, which is equal to,
\begin{equation} \partial_0 R_{i0mj}[h] = -\partial_m R_{i0j0}[h] + \partial_j R_{i0m0}[h], \label{TD1} \end{equation}
by the Bianchi identity.
On the other hand, $R_{i0mj}[h] = \epsilon_{mj}^{\; \; \; \; \; k} {\mathcal B}_{ik}[h]$ and so,
\begin{eqnarray}
\partial_0 R_{i0mj}[h]  &=&  \epsilon_{mj}^{\; \; \; \; \; k} \partial_0 {\mathcal B}_{ik}[h] ,\nonumber \\
&=&  \epsilon_{mj}^{\; \; \; \; \; k} \partial_0 {\mathcal E}_{ik}[f] \label{TD2a}
\end{eqnarray}
by (\ref{EBDuality}).  But $\partial_0 {\mathcal E}_{ik}[f]  = \partial_0 R^m_{\; \; imk}[f] = - \partial^m R_{i0mk}[f]$ by the Bianchi identity and because $R_{0 \; \; \;  \; mk}^{\; \; m}[f] = - R_{0k}[f]$ vanishes on account of step 1.  Hence, 
$\partial_0 {\mathcal E}_{ik}[f]  = - \epsilon_{mk}^{\; \; \; \; \; r} \partial^m {\mathcal B}_{ir}[f]$.  Inserting this expression into (\ref{TD2a}) and using again the twisted self-duality condition yields,
\begin{equation} \partial_0 R_{i0mj}[h] = -\partial_m {\mathcal E}_{ij}[h] + \partial_j {\mathcal E}_{im}[h] .\label{TD2b} \end{equation}
Comparing (\ref{TD1}) with (\ref{TD2b}) gives,
$$
\partial_m \left( R_{i0j0}[h] - {\mathcal E}_{ij}[h]\right)- \partial_j \left(R_{i0m0}[h] - {\mathcal E}_{im}[h] \right)= 0.
$$
Taking into account the definition of  ${\mathcal E}_{ij}[h]$, this is just
\begin{equation}
-\partial_m  \! ^{(4)} \!R_{ij}[h] + \partial_j \! ^{(4)} \! R_{im}[h] = 0. \label{RotR}
\end{equation}
This equation is a bit misleading at first sight, since $\! ^{(4)} \!R_{ij}[h]$ contains $h_{00}$ while neither the electric field nor the magnetic field does, and we started with relations that involved only the electric and the magnetic fields.  But if one writes explicitly $\! ^{(4)} \!R_{ij}[h]$ from its definition in terms of $h_{\mu \nu}$ as
\be 
\! ^{(4)} \!R_{ij}[h] = -\partial_0 K_{ij}[h] + R_{ij}[h] + \frac{1}{2} \partial_i \partial_j h_{00}, \label{DefR}
\ee
with 
$$
 K_{ij}[h] = \frac{1}{2} \left( -\partial_0 h_{ij} + \partial_i h_{j0} + \partial_j h_{i0} \right),
 $$
 (extrinsic curvature)
 and plugs back (\ref{DefR}) in (\ref{RotR}), one sees that $h_{00}$ does drop out from (\ref{RotR}), as it should.  
 
To analyze the implications of the equation (\ref{RotR}), it is easier to write it in terms only of $h_{ij}$ and $h_{0i}$ as
\begin{equation}
\partial_m ( \partial_0 K_{ij}[h] -R_{ij}[h]) - \partial_j (\partial_0 K_{im}[h] - R_{im}[h])= 0. \label{RotPartialK}
\end{equation}
 This equation implies, using the fact that  $\partial_0 K_{ij}[h] $ is symmetric in $(i,j)$
$$
\partial_0 K_{ij}[h] - R_{ij}[h]= \partial_i \partial_j \Phi
$$
for some function  $\Phi$.  Choosing the function $h_{00}$ (which is an arbitrary gauge function not occurring in the original equations (\ref{EBDuality})) to be equal to $2 \Phi$ yields
\be
\! ^{(4)} \!R_{ij}[h] = 0.
\ee
A similar derivation gives 
\be
\! ^{(4)} \!R_{ij}[f] = 0.
\ee
These are the dynamical Einstein equations.  

[Note: The equations obtained above, namely $G_{00}=0$, $R_{0i} \equiv G_{0i}  = 0$ and $\! ^{(4)} \!R_{ij}[h] = 0$ are  the Einstein equations $G_{\al \beta} = 0$ because  $ ^{(4)} \!R[h] = 0$. Indeed, given the constraint $G_{00}=0$ one gets $R_{00} = - \frac{1}{2}  ^{(4)} \!R[h]$ and thus $ \! ^{(4)} \!R[h] = -R_{00} + \! ^{(4)} \!R_{ij}[h] \delta^{ij} = \frac{1}{2}  ^{(4)} \!R[h] +  \! ^{(4)} \!R_{ij}[h] \delta^{ij} \Leftrightarrow  \frac{1}{2}  ^{(4)} \!R[h] =  \! ^{(4)} \!R_{ij}[h] \delta^{ij}$.  This implies $ ^{(4)} \!R[h] = 0$ since $\! ^{(4)} \!R_{ij}[h] = 0$.  Thus one has also $ \! ^{(4)} \!G_{ij}[h] = 0$.]

We thus see that  indeed, the twisted self-duality conditions lead naturally to the dynamical Einstein equations differentiated once,  since equation (\ref{RotR}) contains three derivatives of the metric. But there is no loss of information since one can integrate  (\ref{RotR}) and use the gauge freedom to derive the undifferentiated dynamical Einstein equations.  There is no analog of this feature in the electromagnetic case where the twisted self-duality conditions $F = \! ^* H$, $H =- \! ^* F$, imply the Maxwell equation $d ^* F = 0$, $d ^* H = 0$ ``on the nose".

We can therefore summarize this section by stating that the covariant twisted self-duality conditions (\ref{122}) are entirely equivalent to the spatial twisted self-duality conditions  (\ref{EBDuality}).  This will be used in Section \ref{HETwistedSD} below to establish that the field equations following from (\ref{Action0}) are indeed the twisted self-duality conditions for the two metrics $h_{\mu \nu}$ and its dual $f_{\mu \nu}$.

\section{Gauge invariance of the action (\ref{Action0})}
\label{GaugeInvAction1}
\setcounter{equation}{0}

Before analyzing the field equations following from the action (\ref{Action0}), we first rewrite it in a manner that makes its gauge invariances manifest. To that end, we need some tensor tools from conformal geometry, in particular the concept of (co-)Cotton tensor and its role in three dimensions.  These tools are recalled in the appendices.

As we now show, the action takes a simple form when rewritten in terms of the relevant invariant tensors constructed out of the prepotentials.

\subsection{Rewriting the kinetic term}
Using the expression for the co-Cotton tensor $D^{ij}$ given in the appendix \ref{TensorsLineGrav}, one finds that the kinetic term takes the simple form,
\be
- 2 \int dt \left[ \int d^3x \, \ep^{ab} D_a^{\; ij} \dot{Z}_{bij}  \right], \label{KineticTerm}
\ee
where $D_a^{\; ij} \equiv D^{\; ij}[Z_a]$ is the co-Cotton tensor constructed out of the prepotential $Z_{aij}$ (i.e., replace $h_{ij}$ by $Z_{aij}$ in the formula (\ref{ExprD})).

The invariance of the kinetic term under the transformations (\ref{GaugeTransf})
is obvious after appropriate integrations by parts once it is written in this manner since $D^{\; ij}_a$ itself is not only invariant but also trace-free and divergence-less (see (\ref{IdenD})).

\subsection{Rewriting the Hamiltonian}
The Hamiltonian can also be rewritten in a more transparent manner.  Making appropriate integrations by parts, one gets,
\be
H =\int d^3x \left(4 R^a_{ij} R^{bij} - \frac{3}{2} R^a R^b \right) \de_{ab}, \label{HintermsofR}
\ee
where $R^a_{ij}$ is the Ricci tensor constructed out of the prepotential $Z^a_{ij}$.  Invariance of (\ref{HintermsofR}) under linearized diffeomorphisms is obvious because $R^a_{ij}$ is a tensor.  It is instructive to check also the invariance under conformal transformations. One gets,
\be 
\de H = - 8 \int d^3x G^{aij} \, \partial_{i} \partial_j \ep^b \,  \de_{ab},
\ee
where $G^a_{ij}$ is the Einstein tensor of $Z^a_{ij}$.  Making an integration by parts and using the contracted Bianchi identity yields the desired result,
\be
\de H = 0.
\ee

\subsection{Equations of motion}

Having made the gauge invariance of the action manifest, we now turn to the equations of motion. Their gauge invariance is of course a direct consequence of the gauge invariance of the action, but it is useful to also write them explicitly in terms of invariant tensors as this will facilitate their comparison with the twisted self-duality conditions.

The first-order equations that follow from the action,
\be
S[Z^a_{ij}] = \int dt \left[- 2 \int d^3 \, x \ep^{ab} D^{ij}_a \dot{Z} _{bij} - H \right], \label{Action5}
\ee
are easily found to be
\be
4 \ep^{ab}\dot{D}_a^{ij} + 4 \ep^{imk}\partial_m D^{b j}_{\; \; \; k} = 0 .\label{EOMdiff0}
\ee
The second term is verified to be symmetric in $i,j$.

Now, using the definition of the co-Cotton tensor in terms of the Schouten tensor $S_{ak}^{\; \; \; j}$, 
$$
D^{ij}_a = - \ep^{imk} \partial_m S_{ak}^{\; \; \; j},
$$
one may rewrite these equations as,
\be
\ep^{imk} \partial_m \left[ - \ep^{ab} \dot{S}_{ak}^{\; \; \; j} + D^b_k\,^j \right] = 0 , \label{EOMdiff}
\ee
or equivalently,
\be
\partial_i F^b_{kj} - \partial_k F^b_{ij} = 0, \label{EqForF}
\ee
with 
$F^b_{ij} = - \ep^{ab} \dot{S}_{aij} + D^b_{ij} = F^b_{ji}$. The equation (\ref{EqForF}) implies
$
F^b_{ij} = \partial_i v^b_j
$ for some $v^b_j$ that must be such that $F^b_{ij}$ is symmetric, i.e. $\partial_i v^b_j = \partial_j v^b_i$, which yields in turn $v^b_j = \partial_j \Lambda^b$ for some $\Lambda^b$, giving finally the equation
\be
\ep^{ab} \left(\dot{S}_{aij} - \partial_i \partial_j \Lambda_a \right) =  D^b_{ij} . \label{TwistedSelfDualSup}
\ee

The emergence of $\Lambda_a$ is somewhat similar to the emergence of $A_0$ in the electromagnetic case. The Schouten tensor is not invariant under conformal transformations.  Rather, as shown in the appendix \ref{TensorsLineGrav}, it transforms as
$$
\de S^a_{ij} = - \partial_i \partial_j \epsilon^a.
$$
It is therefore convenient to demand that the $\Lambda^a$ transform as
\be
\de \Lambda^a =  - \dot{\ep}^a,
\ee
so that the quantities
\be
\dot{S}_{aij} - \partial_i \partial_j \Lambda_a \label{SLambda}
\ee
are invariant.  The fields $\Lambda^a$ drop from $\dot{D}^a_{ij}$ and hence also from the action (just as $A_0$ does in the Maxwell case).  We shall see below that (\ref{SLambda}) turns out to be proportional to the magnetic field. 
{}From (\ref{TwistedSelfDualSup}), one can relate $\Lambda^a$ to the trace of $\dot{S}^a_{ij}$ through 
$$
\dot{S}_{aij} \de^{ij}=  \triangle \Lambda_a,
$$
because the co-Cotton tensor $D^b_{ij}$ is traceless.

In the appendix \ref{H1EOMH2EOM}, we compare the above form of the equations of motion to the original Hamiltonian equations derived by varying the action with respect to $h_{ij}$ and $\pi^{ij}$.  This enables one to identify $\Lambda^1$ with the lapse function of $h_{\mu \nu}$ (and by duality, $\Lambda^2$ with the lapse function of the second metric $f_{\mu \nu}$), thus confirming the appropriateness of the analogy with $A_0$.

\section{Hamiltonian equations as twisted self-duality equations}
\label{HETwistedSD}
\setcounter{equation}{0}

We now prove that the equations (\ref{TwistedSelfDualSup}) are just the twisted self-duality equations (\ref{EBDuality}).

To that end, we need to express the electric and magnetic fields of the two metrics in terms of the prepotentials. 

Since we are dealing with the prepotentials, the constraints are identically satisfied.  This means that the electric and magnetic fields are both symmetric and traceless.

\subsection{Electric fields ${\mathcal E}^a_{mn}$}
It follows from formula (\ref{R-D}) in the appendix D that the electric field ${\mathcal E}^a_{mn}$ is equal to
\be
{\mathcal E}^a_{mn} = - 2 D^b_{mn},
\ee
a formula that gives indeed an electric field that is identically symmetric and traceless.

\subsection{Magnetic fields ${\mathcal B}^a_{mn}$}
It is also established in the appendix D that the magnetic fields are related to the time derivatives of the Schouten tensors of the prepotentials as follows,
$$
{\mathcal B}^a_{mn} = 2\dot{ S}^a_{mn} + \partial_m K^a_n,
$$
where the explicit form of $K^a_n $ is given in (\ref{B-S}),
$$
K^a_n = \partial_n \dot{\Phi} - \partial_s \dot{\Phi}^s_{\; \; n} - \ep_{npq}\partial^p(h_0^{\; \; q} - \dot{u}^q).
$$
Because ${\mathcal B}^a_{mn}$ and $\dot{ S}^a_{mn} $ are symmetric, the vector $K^a_n$ is a gradient, and so\footnote{The symmetry of ${\mathcal B}^a_{mn}$ may be viewed as an equation that expresses the shift in terms of the time derivatives of the Schouten tensor.  Indeed, one may decompose the three-dimensional vectors $\vec{K}^a$ as the sum of a gradient plus a curl, $\vec{K}^a = \vec{\nabla} f^a + \vec{\nabla} \times \vec{g}^a$ for some functions $f^a$ and vector fields $\vec{g}^a$.  By adjusting the arbitrary shift vectors $h^a_{0q}$, one can set $g^a = 0$ and the $\vec{K}^a$'s reduce to  the gradients $\vec{\nabla} f^a$.}, 
\be
{\mathcal B}^a_{mn} = 2\dot{ S}^a_{mn} + \partial_m \partial_n f^a,
\ee
for some $f^a$.    As above, the function $f^a$ can be expressed in terms of the trace of $\dot{ S}^a_{mn}$ since ${\mathcal B}^a_{mn}$ is traceless. One gets
$f^a = -2 \Lambda^a$. We have thus demonstrated that the magnetic field is the traceless part of the time derivative of the Schouten tensor.

\subsection{Twisted self-duality conditions}
Combining these results together, we see that the Hamiltonian equations of motion (\ref{TwistedSelfDualSup}) are indeed just the 
twisted self-duality equations (\ref{EBDuality}) since we have the magnetic fields on the left-hand sides and the electric fields on the right-hand sides, with the twisting matrix ${\mathcal S}$ whose matrix elements are $\ep^{ab}$.

\section{Conclusions}
\label{Conclusions}
\setcounter{equation}{0}

In this paper, we have clarified some aspects of the formulation of linearized gravity in terms of prepotentials, which exhibits duality symmetry.  We have shown that the equations of motion can indeed be viewed as twisted self-duality equations, and that the invariance of the action under the gauge symmetries of the prepotentials can also be made explicit.  These results are useful for the prepotential formulation of gravity in higher dimensions \cite{InPrep}, where the dual description involves now a tensor with mixed Young symmetry instead of another symmetric tensor.

We plan to extend also our analysis to higher spin gauge fields which should enjoy similar prepotential formulations \cite{InPrep2}.  Again, we expect the dimension $D=4$ to play a special role in the fully symmetric case \cite{Hull:2001iu,Nieto:1999pn,Bekaert:2003az,Boulanger:2003vs}  since then the two dual descriptions involve tensors of the same types and duality rotations in their internal $2$-space is expected to be a symmetry of the action. In the other dimensions, the dual tensors are of different types but a description of the equations of motion as twisted self-duality conditions is available \cite{InPrep2}.

We now comment on supergravity. It has long been known that the equations of motion for the gravitino may be written as
$$F_{\mu \nu} = \ga_5 ^*\hspace{-.02cm} F_{\mu \nu},$$
where
$$F_{\mu \nu} = \partial_{\mu} \psi_\nu - \partial_\nu \psi_\mu
$$
is the linearized curvature of the spin-$\frac{3}{2}$ field. Therefore $\ga_5$ plays the role of twist matrix in the fermionic sector, and the complete equations of motion of linearized supergravity 
take again the twisted self-duality form 
\begin{equation}
{\mathfrak R} = -{\mathcal S}  \, ^*\hspace{-.02cm}{\mathfrak R}, \label{Sug122}
\end{equation}
with
\begin{equation}
{\mathfrak R} = 
 \begin{pmatrix} R\\ S\\ F \end{pmatrix}, \; \; \; {\mathcal S}  = \begin{pmatrix} 0&-1 & 0\\ 1 & 0 & 0 \\ 0& 0& -\ga_5 \end{pmatrix}  .\label{Sug123}
\end{equation}
We have recently extended \cite{InPrep3} the prepotential formulation of linearized gravity to supergravity. We expect to address in a future publication  \cite{InPrep4} the generalization of the analysis given in this paper for pure linearized gravity to that supersymmetric case.

Finally, it would be of interest to analyze how one can include sources and asymptotic conditions in the twisted self-duality formulation (in this context, see \cite{Bunster:2006rt,Barnich:2008ts,Argurio:2009xr}) and also, to investigate the role of electric-magnetic duality -- which is part of a much bigger group of ``hidden symmetries"  \cite{Cremmer:1978ds} -- at the quantum level, in the line of the work initiated in \cite{Kallosh:2011dp}.

\section*{Acknowledgments} 
C.B. and M.H.  thank  the Alexander von Humboldt Foundation for Humboldt Research Awards.  The work of M.H.and S.H. is partially supported by the ERC through the ``SyDuGraM" Advanced Grant, by IISN - Belgium (conventions 4.4511.06 and 4.4514.08) and by the ``Communaut\'e Fran\c{c}aise de Belgique" through the ARC program.  The Centro de Estudios Cient\'{\i}ficos (CECS) is funded by the Chilean Government through the Centers of Excellence Base Financing Program of Conicyt.   

\vspace{1cm}

\noindent
{\bf \Large{Appendices}}

\appendix
\section{The Cotton tensor}
\label{CottonTensor}
\setcounter{equation}{0}

We recall here well-known elementary concepts from conformal geometry. 
\subsection{Definitions}
We assume the dimension $D \geq 3$.

The Riemann tensor can be decomposed as
$$
R_{\al \bet \ga \de} = W_{\al \bet \ga \de} + \left(S_{\al \ga} g_{\bet \de} - S_{\al \de} g_{\bet \ga} - S_{\bet \ga} g_{\al \de} + S_{\bet \de} g_{\al \ga} \right),
$$
where the ``Weyl tensor" $W_{\al \bet \ga \de}$ fulfills 
\bea
&& W_{\al \bet \ga \de} = - W_{\bet \al \ga \de}, \; \; \; W_{\al \bet \ga \de} = -W_{\al \bet \de \ga}, \; \; \; W_{\al \bet \ga \de} = W_{\ga \de \al \bet }, \nonumber \\
&&  W_{\al \bet \ga \de} + W_{\al \ga \de \bet} + W_{\al \de \bet \ga} = 0, \nonumber 
\eea
and is traceless,
$$
W^\al_{\; \;  \; \bet \al \de} = 0.
$$
The tensor $S_{\al \bet}$ is known as the ``Schouten tensor" and is explicitly given by
$$
S_{\al \bet} = \frac{1}{D-2} \left( R_{\al \bet} - \frac{R}{2(D-1)} g_{\al \bet} \right).
$$

The ``Cotton tensor" $C_{\al \bet \ga}$ is the curl of the Schouten tensor,
$$
C_{\al \bet \ga} = S_{\al \bet ; \ga} - S_{\al \ga ; \bet},
$$
and satisfies
$$
C_{\al \bet \ga}  = - C_{\al \ga \bet} , \; \; \; C_{ \bet \ga \de} + C_{ \ga \de \bet} + C_{ \de \bet \ga} = 0.
$$

\subsection{Bianchi identity}
The Bianchi identity reads
\bea
&& W_{\al \bet \ga \de ; \ep} + W_{\al \bet \de \ep ; \ga} + W_{\al \bet \ep \ga ; \de} + C_{\al \ga \ep} \, g_{\bet \de} + C_{\al \ep \de} \, g_{\bet \ga} +  C_{\al \de \ga} \, g_{\bet \ep} \hspace{2.5cm} \nonumber \\
&& \hspace{5.5cm} -   C_{\bet \ga \ep} \, g_{\al \de} - C_{\bet \ep \de} \, g_{\al \ga} - C_{\bet \de \ga} \, g_{\al \ep}= 0. \label{Bianchi}
\eea
One contraction of the Bianchi identity (\ref{Bianchi}) yields
$$
W^{\al}_{\; \; \; \bet \ga \de ; \al} = (3 - D) C_{\bet \ga \de}.
$$
The contracted Bianchi identity (two contractions of (\ref{Bianchi})) is equivalent to the tracelessness of the Cotton tensor,
$$
C^\al_{\; \; \;  \al \ga} = 0.
$$

\subsection{Conformal transformations}
Under conformal transformations
\be
\hat{g}_{\al \bet} = e^{2 \om} g_{\al \bet}, \label{WeylResc}
\ee
the Weyl, Schouten and Cotton tensors transform respectively as
\bea
&& \hat{W}_{\al \bet \ga \de} = e^{2 \om} W_{\al \bet \ga \de}, \nonumber \\
&& \hat{S}_{\al \bet} = S_{\al \bet} - \om_{\al ; \bet} + \om_\al \om_\bet - \frac{1}{2} g_{\al \bet} \, g^{\la \mu} \om_\la \om_\mu , \nonumber  \\
&& \hat{C}_{\al \bet \ga} = C_{\al \bet \ga} - W^\la_{\; \; \; \al \bet \ga} \om_\la,  \nonumber
\eea
where $\om_\la \equiv \partial_\la \om$.

\section{The case of 3 dimensions}
\label{Case3D}
\setcounter{equation}{0}
\subsection{Conformal invariance of the Cotton tensor}
In three dimensions ($D=3$), the Weyl tensor vanishes identically and the curvature can be expressed as
$$
R_{\al \bet \ga \de} = S_{\al \ga} g_{\bet \de} - S_{\al \de} g_{\bet \ga} - S_{\bet \ga} g_{\al \de} + S_{\bet \de} g_{\al \ga},
$$
in terms of the Schouten tensor,
$$
S_{\al \bet} =  R_{\al \bet} - \frac{R}{4} g_{\al \bet} .
$$

Because the Weyl tensor is identically zero, the Cotton tensor is conformally invariant, i.e., it is inchanged under (\ref{WeylResc}),
\be
\hat{C}_{\al \bet \ga} = C_{\al \bet \ga}.
\ee

\subsection{The traceless, symmetric tensor co-Cotton tensor $D^{\al \bet}$}
It is convenient to introduce the tensor
\be
D^{\al \bet} = \frac{1}{2} \epsilon^{\bet \rh \si} C^\al_{\; \; \; \rh \si}
\ee
equivalent to the Cotton tensor.  We will call it ``the co-Cotton tensor". Because the Cotton tensor is traceless, $D^{\al \bet}$ is symmetric, and because the Cotton tensor fulfills the cyclic identity, $D^{\al \bet}$ is traceless,
$$
D^{\al \bet} = D^{\bet \al}, \; \; \; \; D^\al_{\; \; \;  \al} = 0.
$$
Under conformal transformations,
$$
\hat{D}^{\al \bet} = e^{- 2 \om} D^{\al \bet}.
$$

{}Finally, the tensor $D^{\al \bet}$ is divergence-free,
$$
D^{\al \bet}_{\; \; \; \; \; \; ; \bet} = 0.
$$
In fact, $D^{\al \bet}$ is the tensor that appears in the mass term in the field equations of $D=3$ topologically massive gravity \cite{Deser:1981wh},
$$e
G^{\al \bet} + \frac{1}{\mu} D^{\al \bet} = 0,
$$
and is the functional derivative of the Lorentz Chern-Simons term (with the spin connection treated as the standard function of the triad and its derivatives).

\section{Linearized gravity}
\label{TensorsLineGrav}
\setcounter{equation}{0}
We write the above tensors for linearized gravity, in three dimensions with Euclidean signature (spatial sections of the $3+1$ decomposition).

The Riemann tensor is
$$ R_{ijmn} = \frac{1}{2} \left( \partial_j \partial_m h_{in}  - \partial_i \partial_m h_{jn} - \partial_j \partial_n h_{im} + \partial_i \partial_n h_{jm}\right),
$$
and is invariant under linearized diffeomorphisms,
$$
\delta h_{ij} = \partial_i \xi_j + \partial_j \xi_i.
$$
The Ricci  tensor, the scalar curvature, the Einstein tensor  and the Schouten tensor are respectively given by,
\bea
&& R_{ij} = \frac{1}{2} \left( \partial_i \partial^m h_{mj} + \partial_j \partial^m h_{mi}  - \triangle h_{ij} - \partial_i \partial_j h \right), \nonumber  \\
&& R =  \partial^m \partial^n h_{mn} - \triangle h , \nonumber\\
&& G_{ij} = \frac{1}{2} \left( \partial_i \partial^m h_{mj} + \partial_j \partial^m h_{mi}  - \triangle h_{ij} - \partial_i \partial_j h \right) \nonumber \\
&& \hspace{2cm} - \frac{1}{2} \left( \partial^m \partial^n h_{mn} - \triangle h \right) \delta_{ij} , \label{ExpForG}\\
&& S_{ij} = \frac{1}{2} \left( \partial_i \partial^m h_{mj} + \partial_j \partial^m h_{mi}  - \triangle h_{ij} - \partial_i \partial_j h \right) \nonumber \\
&& \hspace{2cm} - \frac{1}{4} \left( \partial^m \partial^n h_{mn} - \triangle h \right) \delta_{ij}, \nonumber
\eea
where $h \equiv h^m_{\; \; \; m}$. 

Under conformal transformations, 
$$
\delta h_{ij} = 2 \delta_{ij} \ep,
$$
they respectively transform as
$$
\delta R_{ij} = - \partial_i \partial_j \epsilon -  \delta_{ij} \triangle \ep  , \; \;  \delta R = - 4 \triangle \ep, \; \;  \delta G_{ij} = - \partial_i \partial_j \epsilon +  \delta_{ij} \triangle \ep  , \;  \; \delta S_{ij} = - \partial_i \partial_j \epsilon.
$$

The Cotton tensor reads 
\bea
C_{ijk} &=& \frac{1}{2} \left( \partial_i \partial_k \partial^m h_{mj} - \partial_i \partial_j \partial^m h_{mk}\right) + \frac{1}{2} \left(\partial_j \triangle h_{ik} - \partial_k \triangle h_{ij} \right) \nonumber \\
&& - \frac{1}{4} \delta_{ij} \left( \partial_k \partial^n \partial^m h_{mn} - \partial_k \triangle h \right) + \frac{1}{4} \delta_{ik} \left( \partial_j \partial^n \partial^m h_{mn} - \partial_j \triangle h \right) \hspace{.3cm} \nonumber
\eea
and is invariant under conformal transformations,
$$
\delta C_{ijk} = 0.
$$

After some algebra, one finds that the co-Cotton tensor is given by
\be 
D^{ij} = \frac{1}{4} \ep^{imn} \left(\partial^j \partial_n \partial^s h_{sm} - \triangle \partial_n h^j_{\; \;  m} \right) + \frac{1}{4} \ep^{jmn} \left(\partial^i \partial_n \partial^s h_{sm} - \triangle \partial_n h^i_{\; \;  m} \right) .\label{ExprD}
\ee
It is easily verified to be not only symmetric, but also traceless and divergence-free,
\be
D^{ij} = D^{ji}, \; \; \; D^{i}_{\; \; i} = 0, \; \; \; D^{ij}_{\; \; \; \; ,j} = 0. \label{IdenD}
\ee
as it should.

\section{Some useful relations}
\label{UsefulForm}
\setcounter{equation}{0}

It is instructive to write the invariants constructed out of $h_{ij}$ in terms of the invariants constructed out of $Z_{ij}$. 

So, with the definition
\be
h^a_{mn} = \ep_{mrs}\partial^r Z^a_{\; n} +  \ep_{nrs}\partial^r Z^a_{\; m} + \partial_m u^a_n + \partial_n u^a_m ,\label{Defh}
\ee
(see formula (II.11) of \cite{Henneaux:2004jw} with $\Phi \equiv Z^a$), one gets
\be
R_{ij}[h^a] = - 2 D_{ij}[Z^a] . \label{R-D}
\ee
The co-Cotton tensor $D^a_{ij}$ is traceless and conserved. Similarly,  (\ref{Defh}) implies that $R[h^a] = 0$ (it is the general solution of the linearized Hamiltonian constraint) and so $R_{ij}[h^a]$ is also consistently both traceless and conserved (by the Bianchi identity).

Another straightforward computation yields
\be
{\mathcal B}_{mn}[h] = 2 S_{mn}[\dot{\Phi}] + \frac{1}{2} \partial_m \left( \partial_n \dot{\Phi} - \partial_s \dot{\Phi}^s_{\; \; n} - \ep_{npq}\partial^p(h_0^{\; \; q} - \dot{u}^q) \right) .\label{B-S}
\ee
This relation turns out too be key in the identification of the Hamiltonian equations of motion with the twisted self-duality conditions.

Similarly, direct comparison of formula (II.8) of \cite{Henneaux:2004jw} with (\ref{ExpForG}) shows that the conjugate momentum to the metric is given by
\be
\pi^{ij} = 2 G^{ij} [Z^1] , \label{pi-G}
\ee
where $Z^1_{ij} \equiv P_{ij}$ is the first superpotential.  This relation makes it clear that $\pi^{ij}$ is conserved.

Under gauge transformations generated by the Hamiltonian constraint $\int u {\mathcal H}$, the momentum $\pi^{ij}$ transforms as
$$ 
\delta \pi_{ij} = - \partial_i \partial_j u + \de_{ij} \triangle u
$$
This is precisely the transformation of $2 G^{ij} [Z^1]$ provided we identify $u$ with $2 \ep^1$.

\section{More on the equations of motion}
\label{H1EOMH2EOM}
\setcounter{equation}{0}

We verify explicitly here that the equations (\ref{TwistedSelfDualSup}) are equivalent to the Hamiltonian equations in terms of $h_{mn}$ and $\pi^{mn}$.  This is guaranteed of course, but it is instructive to do it.  It also permits to relate the field $\Lambda^a$ to the lapse functions.

The equations of motion in terms of the metric $h_{ij}$ and its conjugate momentum $\pi^{ij}$ read
\begin{eqnarray}
&& \dot{h}_{ij} - \partial_i\xi_j - \partial_j \xi_i = 2\left(\pi_{ij} - \frac{1}{2} \pi \delta_{ij} \right), \nonumber \\
&& \dot{\pi}^{ij} - \partial^i \partial^j u + \delta^{ij} \triangle u = -  G^{ij} [h],  \nonumber
\end{eqnarray}
where $u$ is the lapse.

The second of these equations is easily seen to take the form (\ref{TwistedSelfDualSup}) for $b= 2$.  Indeed, using the expressions (\ref{R-D}) and (\ref{pi-G}) yields
$$
2 G^{ij}[\dot{Z}^1] -  \partial^i \partial^j u + \delta^{ij} \triangle u =  2 D^{ij}[Z^2],
$$
(recalling that $R[h]$ identically vanishes when $h_{ij}$ is viewed as a function of $Z_{ij}^2$ and hence $G^{ij}[h] = R^{ij}[h]$).  Taking the trace of this equation enables one to express $\triangle u$ in terms of $R[\dot{Z}^1]$ and to easily get
$$
S^{ij}[\dot{Z}^1] - \frac{\partial^i \partial^j u}{2} = D^{ij}[Z^2],
$$
as announced.  The derivation also shows that the function $\Lambda^1$ appearing in (\ref{TwistedSelfDualSup}) is the lapse.

To compare the first of the Hamiltonian equations with (\ref{EOMdiff}) with $b=1$, we rewrite it as
\begin{equation}
\dot{h}_{ij} - \partial_i\xi_j - \partial_j \xi_i = 4S_{ij} [Z^1]. \label{interm}
\end{equation}
Now, to get rid of the $\xi_i$-terms, we apply the second-order differential operator defining the Einstein tensor to both sides of (\ref{interm}) to get
$$
-  D_{ij}[\dot{Z}^2] = 2 G_{ij}[S[Z^1]]
$$ 
using(\ref{R-D}).  It is an easy exercise to prove the identity
$$
-\epsilon^{imk}\partial_m D_k^{\; \; j} [Z^1]= 2 G^{ij}[S[Z^1]],
$$
which yields 
$$
  D^{ij}[\dot{Z}^2] = \epsilon^{imk}\partial_m D_k^{\; \; j} [Z^1],
$$ 
i.e., (\ref{EOMdiff0}) with $b=1$.


\begin{thebibliography}{99}

%\cite{Cremmer:1998px}
\bibitem{Cremmer:1998px}
  E.~Cremmer, B.~Julia, H.~Lu and C.~N.~Pope,
  ``Dualisation of dualities. II: Twisted self-duality of doubled fields  and
  superdualities,''
  Nucl.\ Phys.\  B {\bf 535}, 242 (1998)
  [arXiv:hep-th/9806106].
  %%CITATION = NUPHA,B535,242;%%
  
      %\cite{Deser:1976iy}
\bibitem{Deser:1976iy}
  S.~Deser and C.~Teitelboim,
  ``Duality Transformations Of Abelian And Nonabelian Gauge Fields,''
  Phys.\ Rev.\  D {\bf 13}, 1592 (1976).
  %%CITATION = PHRVA,D13,1592;%%
 
 %\cite{Henneaux:1988gg}
\bibitem{Henneaux:1988gg}
  M.~Henneaux and C.~Teitelboim,
  ``Dynamics of Chiral (Self-Dual) $p$-Forms,''
  Phys.\ Lett.\  B {\bf 206}, 650 (1988).
  %%CITATION = PHLTA,B206,650;%%

\bibitem{ScSe} 
  J.~H.~Schwarz and A.~Sen,
  ``Duality symmetric actions,''
  Nucl.\ Phys.\  B {\bf 411}, 35 (1994)
  [arXiv:hep-th/9304154].
  %%CITATION = NUPHA,B411,35;%%

\bibitem{Tonin}
 P.~Pasti, D.~P.~Sorokin and M.~Tonin,
  ``Note on manifest Lorentz and general coordinate invariance in duality
  symmetric models,''
  Phys.\ Lett.\  B {\bf 352}, 59 (1995)
  [arXiv:hep-th/9503182]; \\
  %%CITATION = PHLTA,B352,59;%%
  P.~Pasti, D.~P.~Sorokin and M.~Tonin,
  ``Duality symmetric actions with manifest space-time symmetries,''
  Phys.\ Rev.\  D {\bf 52}, 4277 (1995)
  [arXiv:hep-th/9506109]; \\
  %%CITATION = PHRVA,D52,4277;%%
  I.~A.~Bandos, N.~Berkovits and D.~P.~Sorokin,
  ``Duality-symmetric eleven-dimensional supergravity and its coupling to
  M-branes,''
  Nucl.\ Phys.\  B {\bf 522}, 214 (1998)
  [arXiv:hep-th/9711055].
  %%CITATION = NUPHA,B522,214;%%
  
    %\cite{DGHT}
\bibitem{DGHT}
  S.~Deser, A.~Gomberoff, M.~Henneaux and C.~Teitelboim,
  ``Duality, self-duality, sources and charge quantization in abelian  N-form
  theories,''
  Phys.\ Lett.\  B {\bf 400}, 80 (1997)
  [arXiv:hep-th/9702184]; \\
  %%CITATION = PHLTA,B400,80;%%
  S.~Deser, A.~Gomberoff, M.~Henneaux and C.~Teitelboim,
  ``P-brane dyons and electric magnetic duality,''
  Nucl.\ Phys.\ B {\bf 520}, 179 (1998)
  [hep-th/9712189].
  %%CITATION = HEP-TH/9712189;%%

  %\cite{Bunster:2011aw}
\bibitem{Bunster:2011aw} 
  C.~Bunster and M.~Henneaux,
  ``Sp(2n,R) electric-magnetic duality as off-shell symmetry of interacting electromagnetic and scalar fields,''
  PoS HRMS {\bf 2010}, 028 (2010)
  [arXiv:1101.6064 [hep-th]].
  %%CITATION = ARXIV:1101.6064;%%

 %\cite{Bunster:2011qp}
\bibitem{Bunster:2011qp} 
  C.~Bunster and M.~Henneaux,
  ``The Action for Twisted Self-Duality,''
  Phys.\ Rev.\ D {\bf 83}, 125015 (2011)
  [arXiv:1103.3621 [hep-th]].
  %%CITATION = ARXIV:1103.3621;%%
  
 %\cite{Curtright:1980yk}
\bibitem{Curtright:1980yk} 
  T.~Curtright,
  ``Generalized Gauge Fields,''
  Phys.\ Lett.\ B {\bf 165}, 304 (1985).
  %%CITATION = PHLTA,B165,304;%%
   %\cite{Hull:2001iu}

\bibitem{Hull:2001iu} 
  C.~M.~Hull,
  ``Duality in gravity and higher spin gauge fields,''
  JHEP {\bf 0109}, 027 (2001)
  [hep-th/0107149].
  %%CITATION = HEP-TH/0107149;%%
    
 \bibitem{InPrep} C. Bunster, M. Henneaux, S. H\"ortner, ``Twisted self-duality for gravity in abitrary dimensions", in preparation.
 
 %\cite{Henneaux:2004jw}
\bibitem{Henneaux:2004jw}
  M.~Henneaux and C.~Teitelboim,
  ``Duality in linearized gravity,''
  Phys.\ Rev.\  D {\bf 71}, 024018 (2005)
  [arXiv:gr-qc/0408101].
  %%CITATION = PHRVA,D71,024018;%%
 
 %\cite{Julia:2005ze}
\bibitem{Julia:2005ze} 
  B.~Julia, J.~Levie and S.~Ray,
  ``Gravitational duality near de Sitter space,''
  JHEP {\bf 0511}, 025 (2005)
  [hep-th/0507262].
  %%CITATION = HEP-TH/0507262;%%

  %\cite{Nieto:1999pn}
\bibitem{Nieto:1999pn} 
  J.~A.~Nieto,
  ``S duality for linearized gravity,''
  Phys.\ Lett.\ A {\bf 262}, 274 (1999)
  [hep-th/9910049].
  %%CITATION = HEP-TH/9910049;%%

 %\cite{Deser:2004xt}
\bibitem{Deser:2004xt} 
  S.~Deser and D.~Seminara,
  ``Duality invariance of all free bosonic and fermionic gauge fields,''
  Phys.\ Lett.\ B {\bf 607}, 317 (2005)
  [hep-th/0411169].
  %%CITATION = HEP-TH/0411169;%%

\bibitem{InPrep2}
C. Bunster, M. Henneaux, S. H\"ortner, ``Twisted self-duality for higher spin gauge fields", in preparation.

  %\cite{Bekaert:2003az}
\bibitem{Bekaert:2003az} 
  X.~Bekaert and N.~Boulanger,
  ``On geometric equations and duality for free higher spins,''
  Phys.\ Lett.\ B {\bf 561}, 183 (2003)
  [hep-th/0301243].
  %%CITATION = HEP-TH/0301243;%%


  %\cite{Boulanger:2003vs}
\bibitem{Boulanger:2003vs} 
  N.~Boulanger, S.~Cnockaert and M.~Henneaux,
  ``A note on spin s duality,''
  JHEP {\bf 0306}, 060 (2003)
  [hep-th/0306023].
  %%CITATION = HEP-TH/0306023;%%
  
 

\bibitem{InPrep3}
C. Bunster, M. Henneaux, ``Supersymmetric electric-magnetic duality as a manifest symmetry of the action for 
super-Maxwell theory and linearized supergravity", to be published.

\bibitem{InPrep4}
C. Bunster, M. Henneaux, S. H\"ortner, ``Electric-Magnetic Duality in Supergravity, Gauge Invariance and Twisted Self-Duality", in preparation.

 %\cite{Bunster:2006rt}
\bibitem{Bunster:2006rt} 
  C.~W.~Bunster, S.~Cnockaert, M.~Henneaux and R.~Portugues,
  ``Monopoles for gravitation and for higher spin fields,''
  Phys.\ Rev.\ D {\bf 73}, 105014 (2006)
  [hep-th/0601222].
  %%CITATION = HEP-TH/0601222;%%

  %\cite{Barnich:2008ts}
\bibitem{Barnich:2008ts} 
  G.~Barnich and C.~Troessaert,
  ``Manifest spin 2 duality with electric and magnetic sources,''
  JHEP {\bf 0901}, 030 (2009)
  [arXiv:0812.0552 [hep-th]].
  %%CITATION = ARXIV:0812.0552;%%

   %\cite{Argurio:2009xr}
\bibitem{Argurio:2009xr} 
  R.~Argurio and F.~Dehouck,
  ``Gravitational duality and rotating solutions,''
  Phys.\ Rev.\ D {\bf 81}, 064010 (2010)
  [arXiv:0909.0542 [hep-th]].
  %%CITATION = ARXIV:0909.0542;%%
  
 %\cite{Cremmer:1978ds}
\bibitem{Cremmer:1978ds} 
  E.~Cremmer and B.~Julia,
  ``The N=8 Supergravity Theory. 1. The Lagrangian,''
  Phys.\ Lett.\ B {\bf 80}, 48 (1978).
  %%CITATION = PHLTA,B80,48;%%

  
  %\cite{Kallosh:2011dp}
\bibitem{Kallosh:2011dp} 
  R.~Kallosh,
  ``$E_{7(7)}$ Symmetry and Finiteness of N=8 Supergravity,''
  JHEP {\bf 1203}, 083 (2012)
  [arXiv:1103.4115 [hep-th]].
  %%CITATION = ARXIV:1103.4115;%%
  
 %\cite{Deser:1981wh}
\bibitem{Deser:1981wh}
  S.~Deser, R.~Jackiw and S.~Templeton,
  ``Topologically Massive Gauge Theories,''
  Annals Phys.\  {\bf 140}, 372 (1982)
  [Erratum-ibid.\  {\bf 185}, 406 (1988)]
  [Annals Phys.\  {\bf 185}, 406 (1988)]
  [Annals Phys.\  {\bf 281}, 409 (2000)].
  %%CITATION = APNYA,281,409;%%
 
  
  
\end{thebibliography}
\end{document}